\documentclass[Conference, 10pt]{IEEEtran}
\IEEEoverridecommandlockouts
\usepackage{times}
\usepackage{cite}
\usepackage{amsmath,amssymb,amsfonts}
\usepackage{algorithmic}
\usepackage{graphicx}
\usepackage{textcomp}
\usepackage{xcolor}
\usepackage{algorithm}
\usepackage{array}
\usepackage{stfloats}
\usepackage{url}
\usepackage{verbatim}
\usepackage{graphics}
\usepackage{color} 
\usepackage{xpatch}
\usepackage{epstopdf}
\usepackage{epsfig}
\usepackage{cuted}
\usepackage{subfigure} 
\def\BibTeX{{\rm B\kern-.05em{\sc i\kern-.025em b}\kern-.08em
    T\kern-.1667em\lower.7ex\hbox{E}\kern-.125emX}}
\begin{document}

\title{Receiver Design for Continuous-Time AFDM based on  Pre-Filter De-Chirp\vspace{0.5em}\\
\thanks{This work was supported in part by the Natural Science Foundation of China under Grant 62671119; in part by the Natural Science Foundation of Sichuan Province under Grant 2026NSFSC0399; in part by Hainan Province Science and Technology Plan Projects under Grants ZDYF2026LALH001 and ZDYF2025(LALH)001.}
}
\author{\IEEEauthorblockN{Qingyu~Li$^1$, Shuangyang~Li$^2$, Guanghui~Liu$^1$, Fuchen~Xu$^1$, Hongjun~Liu$^1$ } \\
	
	\IEEEauthorblockA{$^1$School of Information and Communication Engineering,
		University of Electronic Science and Technology of China (UESTC), Chengdu
		611731, China\\
		Emails:  \{qingyu.li@std., guanghuiliu@, hjliu@std., fuchenxu@std.\}uestc.edu.cn}
		
	\IEEEauthorblockA{$^2$Faculty of Electrical Engineering and Computer Science, Technical University of Berlin, Berlin 10587, Germany\\
	Emails: shuangyang.li@tu-berlin.de}\vspace{-1em}}

\maketitle
\thispagestyle{empty}
\begin{abstract}
Affine frequency division multiplexing (AFDM) is a promising waveform for future high-mobility communications due to its Doppler robustness and diversity gain. However, existing continuous-time AFDM relies on chirp-dependent wideband receive filtering, which increases noise power and causes channel-energy leakage. In this paper, we propose an AFDM reception scheme based on pre-filter de-chirp for doubly selective channels. The coupling between de-chirp and the channel is first characterized, yielding an equivalent path representation with delay-coupled effective Doppler shifts. The discrete affine Fourier transform (DAFT)-domain equivalent channel is then derived under general transmit pulse shaping and linear time-invariant receive filtering, showing that the effective filtering kernel depends only on the relative sample offset and avoids the additional spreading caused by conventional chirp-dependent filtering. The receive-noise power is further analyzed, and an explicit expression for the signal-to-noise ratio (SNR) gap is obtained by the required receive-filter bandwidths. Simulations verify the theoretical SNR analysis and show that the proposed scheme significantly outperforms the conventional scheme while approaching ideal discrete-time AFDM.
\end{abstract}


\section{Introduction}

Orthogonal frequency division multiplexing (OFDM) has been widely adopted in current cellular networks, but it may suffer severe
performance degradation in highly dynamic channels due to its limited
robustness against large Doppler shifts \cite{11373535}. Orthogonal time
frequency space (OTFS) modulation improves Doppler resilience by
representing the channel in the delay--Doppler (DD) domain
\cite{li2026otfs,22,9404861}. However, its two-dimensional frame structure increases
processing complexity and latency \cite{9508932}, while fractional Doppler
shifts may cause DD-domain spreading, degrading channel sparsity and
increasing pilot overhead \cite{44,33}.

Recently, affine frequency division multiplexing (AFDM) \cite{99}, based on
the discrete affine Fourier transform (DAFT), has attracted considerable
attention. By employing chirp-based subcarriers, AFDM provides enhanced
robustness to Doppler variations \cite{10845819} and can achieve full
diversity over doubly selective channels with a one-dimensional signaling
structure \cite{11263921,10769778}. Moreover, it can preserve a relatively
sparse equivalent channel under fractional Doppler shifts, potentially
reducing pilot overhead compared with OTFS \cite{11557489}.

Continuous-time (CT) modeling is essential for practical waveform design,
as pulse shaping and receive filtering cannot be fully captured by
discrete-time models \cite{raviteja2019practical}. However, most AFDM
studies remain discrete-time. Mirabella \emph{et al.}
\cite{mirabella2026continuous} first studied CT AFDM pulse shaping and
receive filtering. Their framework requires subcarrier suppression (SC),
causing spectral-efficiency loss, and chirp-dependent wideband receive
filtering, which increases noise bandwidth and receive-noise power.
Practical finite-duration filters may further introduce channel spreading
and interference, motivating the pre-filter de-chirp scheme proposed in
this work.

In this paper, we propose an AFDM reception scheme based on pre-filter
de-chirp for doubly selective channels. A unified analytical framework is
developed to characterize the proposed scheme and evaluate its performance
from both channel and noise perspectives. 

The main
contributions are summarized as follows:

\begin{itemize}
	
	\item We propose a pre-filter de-chirp scheme for CT AFDM and
	characterize its coupling with doubly selective channels. The resulting
	signal model shows that each propagation path can be equivalently described
	by an equivalent gain, delay, and delay-coupled effective Doppler shift, which enables
	linear time-invariant (LTI) receive filtering after the de-chirp operation.
	
\item We derive the DAFT-domain equivalent channel of the proposed scheme
under general transmit pulse shaping and LTI receive filtering, without
relying on a specific pulse-shaping or frequency-sampling representation.
The resulting formulation shows that the effective filtering kernel depends
only on the relative sample offset, avoiding the additional
sampling-dependent weighting of the conventional scheme.
	
	\item We analyze the receive-noise power of the proposed de-chirp-based scheme and the conventional
	scheme and derive an explicit expression for their SNR gap in terms of the
	required receive-filter bandwidths. Simulation results verify the derived
	SNR gap.
	
\end{itemize}

The remainder of this paper is organized as follows. Section II presents the CT AFDM baseband system model. Section III introduces the signal model and equivalent channel of the proposed de-chirp scheme. Section IV analyzes the receive-noise power. Section V presents simulation results for performance validation. Section VI concludes the paper.

\section{Conventional Continuous-Time AFDM}
This section presents the CT baseband model of the
conventional AFDM scheme in \cite{mirabella2026continuous}, focusing on
the transmit pulse-shaping  and the chirp-dependent receive filter.

An AFDM block consisting of $N$ time-domain samples is generated through 
the inverse discrete affine Fourier transform (IDAFT), which is given by
$
	x[k]
	=
	\frac{1}{\sqrt{N}}
	\sum_{q\in\mathcal{Q}}
	d_q
	e^{j2\pi\left(
		c_1 k^2+c_2 q^2+\frac{qk}{N}
		\right)},
	\quad k=0,\ldots,N-1,
	\label{eq:afdm_discrete_tx}
$
where $N$ denotes the number of AFDM subcarriers; $d_q$ is the data
symbol transmitted on the $q$-th subcarrier; and $c_1$ and $c_2$ are
the two chirp parameters of the DAFT. Throughout this paper, centered
DAFT-domain indexing is adopted, i.e.,
$q,m\in\{-N/2,\ldots,N/2-1\}$, and the DAFT chirp factors are defined
with respect to these centered indices. The set
$\mathcal{Q}\subseteq\{-N/2,\ldots,N/2-1\}$ denotes the set of active
subcarriers.

Let $T_s$ denote the sampling interval, which is the inverse of the
total bandwidth, and $p(t)$ denote the transmit
pulse-shaping waveform. We further define the CT chirp
parameter as $\bar{c}_1\triangleq c_1/T_s^2$. The corresponding
CT complex AFDM signal within one AFDM block can
then be expressed as
\begin{equation}
	s(t)=
	\sum_{k=0}^{N-1}
	x[k]p(t-kT_s)
	e^{-j2\pi c_1k^2}
	e^{j2\pi\bar{c}_1t^2},
	\label{eq:continuous_afdm_tx}
\end{equation}
where $k$ denotes the time-domain sample index and $s(t)$ denotes the 
CT complex baseband transmit signal.

In the conventional CT AFDM scheme \cite{mirabella2026continuous}, the transmit pulse and 
active subcarriers are properly designed to avoid the superposition of 
spectral replicas and to maintain an approximately uniform gain over the 
active subcarriers. In practice, a root-raised-cosine (RRC) pulse together 
with SCs is adopted to satisfy these waveform requirements. Under these
conditions, the $q$-th active-subcarrier component admits the simplified
CT representation \cite{mirabella2026continuous}
\begin{equation}
	s_{{\rm A},q}(t)
	=\frac{d_qP_q}{\sqrt N T_s}
	e^{j2\pi c_2q^2}
	e^{j2\pi q\Delta f t}
	e^{j2\pi\bar c_1 t^2},
\end{equation}
where $P_q=P(q\Delta f)$ and $P(f)$ is the Fourier transform of $p(t)$.

We consider a doubly selective channel consisting of $L$ propagation 
paths. The time-varying channel impulse response is modeled as
\begin{equation}
	h(t,\tau)
	=
	\sum_{\ell=0}^{L-1}
	h_{\ell} e^{j2\pi \nu_{\ell} t}
	\delta(\tau-\tau_{\ell}),
	\label{eq:channel_model}
\end{equation}
where $h_{\ell}$, $\tau_{\ell}$, and $\nu_{\ell}$ denote the complex 
channel coefficient, propagation delay, and Doppler shift of the $\ell$-th path, 
respectively, and $\delta(\cdot)$ denotes the Dirac delta function.

Accordingly, the received CT baseband signal is written as
\begin{align}
	r(t)
	&=
	\sum_{\ell=0}^{L-1}
	h_{\ell}s(t-\tau_{\ell})
	e^{j2\pi \nu_{\ell}t}
	+w(t),
	\label{eq:received_signal}
\end{align}
where $w(t)$ denotes the complex additive white Gaussian noise (AWGN).

Let $t_0$ denote the global reference time of the receive observation window. In \cite{mirabella2026continuous}, the received signal $r(t)$ is processed by a 
chirp-dependent filter given by
\begin{equation}
	g_{\rm A}(t)
	=
	u_{\rm A}^{*}(-t)e^{-j2\pi \bar{c}_1t^2},
	\label{eq:conv_rx_filter}
\end{equation}
where $u_{\rm A}(t)$ is the underlying receive-filter impulse response; 
$(\cdot)^*$ denotes conjugation. The filtered signal is expressed as
$
	r_{\rm A}(t)=r(t)*g_{\rm A}(t),
$
where $*$ denotes linear convolution.

Under the transmit-waveform conditions in
\cite{mirabella2026continuous}, the received component corresponding to
the $q$-th active subcarrier and the $\ell$-th propagation path can be
written as
\begin{equation}
	r_{q,\ell}(t)
	=\frac{h_\ell d_qP_q}{\sqrt N T_s}
	e^{j2\pi c_2q^2}
	e^{j2\pi[
		\bar c_1(t-\tau_\ell)^2
		+q\Delta f(t-\tau_\ell)
		+\nu_\ell t]}.
	\label{eq:conv_subcarrier_component}
\end{equation}

After passing through the chirp-dependent receive filter
$g_{\rm A}(t)$, its contribution is given by
\begin{align}
	(r_{q,\ell}*g_{\rm A})(t)
	&=\frac{h_\ell d_qP_q}{\sqrt N T_s}
	e^{j2\pi[
		\bar{c}_1(t-\tau_\ell)^2
		+q\Delta f(t-\tau_\ell)
		+\nu_\ell t]}
	\nonumber\\
	&\quad\times
	U_{\rm A}^{*}
	\left(
	q\Delta f+\nu_\ell
	+2\bar{c}_1(t-\tau_\ell)
	\right),
	\label{eq:conv_filter_derivation}
\end{align}
where $U_{\rm A}(f)$ denotes the Fourier transform of
$u_{\rm A}(t)$, and $\Delta f=1/(NT_s)$ is the subcarrier spacing.

Sampling the filtered signal at
$t=t_0+nT_s$ yields
\begin{align}
	r_{\rm A}[n]
	&=
	r_{\rm A}(t)\big|_{t=t_0+nT_s}
	\nonumber=
	\frac{1}{\sqrt N}\sum_{\ell=0}^{L-1}
	\sum_{q\in\mathcal Q}
	h_\ell d_q P_q
	e^{j2\pi c_2q^2}
	\nonumber\\
	&\quad\times
	e^{j2\pi[
		\bar c_1(t_0+nT_s-\tau_\ell)^2
		+q\Delta f(t_0+nT_s-\tau_\ell)
		+\nu_\ell(t_0+nT_s)]}
	\nonumber\\
	&\quad\times
	U_{\rm A}^{*}
	\left(
	q\Delta f+\nu_\ell
	-2\bar c_1\tau_\ell
	+2\bar c_1(t_0+nT_s)
	\right)
	+w_{\rm A}[n].
	\label{eq:conv_sampled_signal}
\end{align}

The DAFT-domain received symbol is obtained by applying the AFDM
demodulation to $r_{\rm A}[n]$, i.e.,
$
	y[m]
	=
	\frac{1}{\sqrt{N}}
	\sum_{n=0}^{N-1}
	r_{\rm A}[n]
	e^{-j2\pi
		\left(
		c_1n^2+c_2m^2+\frac{mn}{N}
		\right)}.
	\label{eq:conv_daft_demod}
$

It can be observed from \eqref{eq:conv_sampled_signal} that the
frequency argument of the receive filter varies with the sampling index
$n$. Consequently, any nonuniformity of the receive-filter response over
the chirp-induced frequency sweep results in sampling-dependent weighting,
which may lead to DAFT-domain energy spreading and additional
inter-carrier interference (ICI). Therefore, the conventional receive
filter requires a sufficiently wide and flat frequency response to
maintain an approximately constant gain over the desired signal
components throughout an AFDM block.

\section{De-Chirp AFDM}
In this section, we first characterize the proposed de-chirp AFDM scheme
over doubly selective channels under general transmit pulse shaping and
LTI receive filtering. We then derive its DAFT-domain equivalent channel
without specifying particular pulse shapes or LTI filter responses, and
compare it with that of the conventional scheme.

\subsection{Received Signal Model}

\begin{figure}
	\centering
	\includegraphics[width=0.49\textwidth]{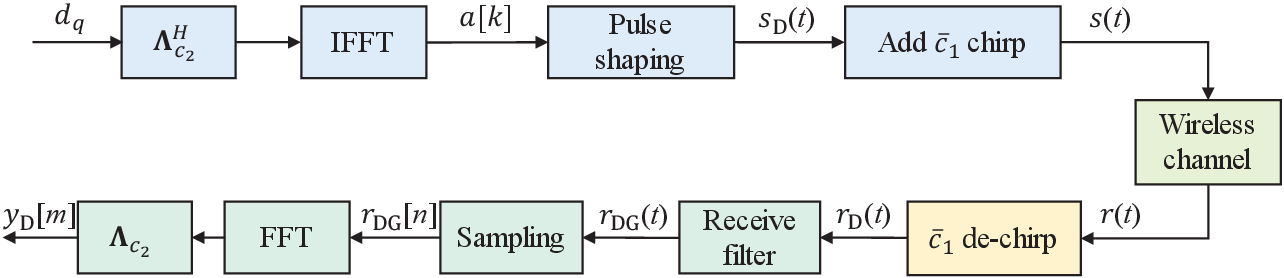}
	\caption{Block diagram of the proposed de-chirp AFDM system.}
	\label{str}
\end{figure}

As shown in Fig.~\ref{str}, unlike
\cite{mirabella2026continuous}, where $r(t)$ in
\eqref{eq:received_signal} is directly passed through a wideband receive
filter, the proposed scheme first performs de-chirp on $r(t)$ with
respect to the same global reference time $t_0$, i.e.,
\begin{equation}
	r_{\rm D}(t)
	=
	r(t)e^{-j2\pi\bar{c}_1(t-t_0)^2}.
	\label{eq:dechirp_signal}
\end{equation}
The de-chirp operation removes the quadratic time-varying
phase and yields the equivalent path representation
derived below.

Considering the contribution of the $\ell$-th propagation path, the
de-chirped received signal can be expressed as
\begin{align}
	r_{{\rm D},\ell}(t)
	&=
	h_\ell e^{j2\pi\nu_\ell t}
	e^{j2\pi\bar{c}_1(t-\tau_\ell)^2}
	e^{-j2\pi\bar{c}_1(t-t_0)^2}
	s_{\rm D}(t-\tau_\ell)
	\nonumber\\
	&=
	h_\ell
	e^{j2\pi\bar{c}_1(\tau_\ell^2-t_0^2)}
	e^{j2\pi[\nu_\ell+2\bar{c}_1(t_0-\tau_\ell)]t}
	s_{\rm D}(t-\tau_\ell),
	\label{eq:dechirp_path}
\end{align}
where
$
	s_{\rm D}(t)
	\triangleq
	s(t)e^{-j2\pi\bar{c}_1t^2}.
	\label{eq:sD_definition}
$

We define the time-invariant equivalent path gain as
\begin{equation}
	\tilde{h}_\ell
	\triangleq
	h_\ell
	e^{j2\pi\bar{c}_1(\tau_\ell^2-t_0^2)},
	\label{eq:equiv_gain}
\end{equation}
and the corresponding effective Doppler shift as
\begin{equation}
	\mu_\ell
	\triangleq
	\nu_\ell+2\bar{c}_1(t_0-\tau_\ell).
	\label{eq:equiv_doppler}
\end{equation}

To retain a general pulse-shaping model, define
$
	a[k]
	\triangleq
	x[k]e^{-j2\pi c_1k^2}.
	\label{eq:ak_general}
$
According to \eqref{eq:continuous_afdm_tx},
$
	s_{\rm D}(t)
	=
	\sum_{k=0}^{N-1}
	a[k]p(t-kT_s),
	\label{eq:sD_general}
$
where
\begin{equation}
	a[k]
	=
	\frac{1}{\sqrt{N}}
	\sum_{q\in\mathcal Q}
	d_q
	e^{j2\pi\left(c_2q^2+\frac{qk}{N}\right)}.
	\label{eq:ak_data}
\end{equation}
Hence, \eqref{eq:dechirp_path} can be written  as
\begin{equation}
	\begin{aligned}
	r_{{\rm D},\ell}(t)
	&=
	\tilde{h}_{\ell} e^{j2\pi\mu_{\ell} t}
	s_{\mathrm{D}}(t-\tau_{\ell})\\&=
	\tilde h_\ell e^{j2\pi\mu_\ell t}
	\sum_{k=0}^{N-1}
	a[k]p(t-\tau_\ell-kT_s).
	\label{eq:dechirp_path_general}
\end{aligned}
\end{equation}

Equation \eqref{eq:dechirp_path_general} shows that, after de-chirp, the
quadratic time-varying phase associated with the $c_1$ chirp is removed.
Each propagation path is therefore represented by a delayed
pulse-shaped waveform multiplied by a single complex exponential with the
delay-coupled effective Doppler shift $\mu_\ell$.

After de-chirp, a conventional LTI receive filter can be employed. Let
$g_{\rm D}(t)$ denote its impulse response. According to
\eqref{eq:dechirp_path_general}, the contribution of the $\ell$-th path
after receive filtering is expressed as
\begin{align}
	r_{{\rm DG},\ell}(t)
	&=
	\int_{-\infty}^{\infty}
	r_{{\rm D},\ell}(t-\lambda)
	g_{\rm D}(\lambda)\,{\rm d}\lambda
	\nonumber\\
	&=
	\tilde h_\ell e^{j2\pi\mu_\ell t}
	\sum_{k=0}^{N-1}a[k]
	\int_{-\infty}^{\infty}
	g_{\rm D}(\lambda)e^{-j2\pi\mu_\ell\lambda}\nonumber
	\\&\qquad\times p(t-\tau_\ell-kT_s-\lambda)\,{\rm d}\lambda .
	\label{eq:dechirp_filter_expansion}
\end{align}
	
For notational convenience, we define the corresponding effective
filtered pulse as
\begin{align}
	\psi_\ell(t)
	=&
	p(t)*
	\left[
	g_{\rm D}(t)e^{-j2\pi\mu_\ell t}
	\right]\nonumber\\	\triangleq&
	\int_{-\infty}^{\infty}
	g_{\rm D}(\lambda)
	e^{-j2\pi\mu_\ell\lambda}
	p(t-\lambda)
	{\rm d}\lambda
	\label{eq:effective_filtered_pulse}.
\end{align}

Therefore, the filtered CT received signal is \begin{align} r_{\rm DG}(t) = \sum_{\ell=0}^{L-1} \tilde h_\ell e^{j2\pi\mu_\ell t} \sum_{k=0}^{N-1} a[k] \psi_\ell(t-\tau_\ell-kT_s) +w_{\rm D}(t), \label{eq:dechirp_filtered_signal} \end{align} where $w_{\rm D}(t)$ denotes the noise after de-chirp and filtering.  In general, the filtering operation may introduce noise
correlation. However, for practical filters with rapidly decaying noise
autocorrelation, the resulting covariance matrix is often approximately
diagonal.

The filtered signal is then sampled at
\begin{equation}
	t=t_0+nT_s,
	\qquad n=0,\ldots,N-1.
\end{equation}
Let
$
	T=NT_s={1}/{\Delta f},
	\kappa_\ell
	\triangleq
	\mu_\ell T,
	\label{eq:normalized_parameters}
$
and absorb the time-independent phase associated with $t_0$ into
\begin{equation}
	\breve h_\ell
	\triangleq
	\tilde h_\ell e^{j2\pi\mu_\ell t_0}.
	\label{eq:effective_gain_sampling}
\end{equation}
The sampled signal can then be expressed as
\begin{align}
	r_{\rm DG}[n]
	&=
	\sum_{\ell=0}^{L-1}
	\breve h_\ell
	e^{j2\pi\kappa_\ell n/N}
	\sum_{k=0}^{N-1}
	a[k]
	\psi_\ell
	\left(
	t_0-\tau_\ell+(n-k)T_s
	\right)
	\nonumber\\
	&\quad
	+w_{\rm D}[n].
	\label{eq:dechirp_sampled_general}
\end{align}

A key observation from \eqref{eq:dechirp_sampled_general} is that, for
each propagation path, the receive filtering is characterized by the
fixed effective pulse $\psi_\ell(t)$ and does not introduce an additional
sampling-index-dependent filter response. This differs fundamentally from
the chirp-dependent receive filtering in the conventional scheme \cite{mirabella2026continuous}.

Since the $c_1$-dependent quadratic phase has already been removed by
the de-chirp operation, the received samples are transformed into the
DAFT domain by
\begin{equation}
	y_{\rm D}[m]
	=
	e^{-j2\pi c_2m^2}
	\frac{1}{\sqrt{N}}
	\sum_{n=0}^{N-1}
	r_{\rm DG}[n]e^{-j2\pi mn/N}.
	\label{eq:dechirp_daft}
\end{equation}
Substituting \eqref{eq:dechirp_sampled_general} into
\eqref{eq:dechirp_daft} gives
\begin{align}
	y_{\rm D}[m]
	&=
	e^{-j2\pi c_2m^2}
	\frac{1}{\sqrt N}
	\sum_{\ell=0}^{L-1}
	\breve h_\ell
	\sum_{n=0}^{N-1}
	e^{-j2\pi(m-\kappa_\ell)n/N}
	\nonumber\\
	&\quad\times
	\sum_{k=0}^{N-1}
	a[k]
	\psi_\ell
	\left(
	t_0-\tau_\ell+(n-k)T_s
	\right)
	+v_{\rm D}[m],
	\label{eq:dechirp_daft_io}
\end{align}
where $v_{\rm D}[m]$ denotes the corresponding noise component in the
DAFT domain.

From a practical implementation perspective, the proposed receive filter
does not require the knowledge of individual delay and Doppler parameters of all
propagation paths. Its passband can be designed to cover the occupied
spectrum of $s_{\rm D}(t)$ shifted by the possible range of effective
Doppler shifts determined from the maximum channel delay and Doppler
spread.

\subsection{DAFT-Domain Equivalent Channel}

To obtain a compact DAFT-domain representation, define the unitary
$N$-point DFT matrix $\mathbf F$ as
\begin{equation}
	[\mathbf F]_{m,n}
	\triangleq
	\frac{1}{\sqrt N}
	e^{-j2\pi mn/N}.
	\label{eq:DFT_matrix}
\end{equation}
and define the $N\times N$ diagonal ${c_2}$-chirp matrix
\begin{equation}
	[\mathbf{\Lambda}_{c_2}]_{m,m}
	\triangleq
	e^{-j2\pi c_2m^2}.
	\label{eq:c2_matrix}
\end{equation}
Let $\mathbf d\in\mathbb C^N$ denote the DAFT-domain symbol vector. From \eqref{eq:ak_data}, the
$c_1$-de-chirped time-domain sequence satisfies
\begin{equation}
	\mathbf a
	=
	\mathbf F^{H}
	\mathbf{\Lambda}_{c_2}^{H}
	\mathbf d.
	\label{eq:a_matrix}
\end{equation}

For the $\ell$-th path, define the $N\times N$ diagonal matrix
\begin{equation}
	\left[\mathbf D(\kappa_\ell)\right]_{n,n}
	\triangleq
	e^{j2\pi\kappa_\ell n/N},
	\quad n=0,\ldots,N-1 .
	\label{eq:doppler_diag}
\end{equation}
and the sampled pulse-filtering matrix
\begin{equation}
	\left[
	\mathbf{\Phi}_{{\rm D},\ell}
	\right]_{n,k}
	\triangleq
	\psi_\ell
	\left(
	t_0-\tau_\ell+(n-k)T_s
	\right),
	\label{eq:dechirp_pulse_matrix}
\end{equation}
where $\mathbf{\Phi}_{D,\ell}\in\mathbb C^{N\times N}$.
Then \eqref{eq:dechirp_sampled_general} can be written in vector form as
\begin{equation}
	\mathbf r_{\rm DG}
	=
	\sum_{\ell=0}^{L-1}
	\breve h_\ell
	\mathbf D(\kappa_\ell)
	\mathbf{\Phi}_{{\rm D},\ell}
	\mathbf a
	+
	\mathbf v_{\rm DG}.
	\label{eq:dechirp_sampled_matrix}
\end{equation}
Consequently,
\begin{equation}
	\mathbf y_{\rm D}
	=
	\mathbf H_{\rm D}\mathbf d
	+
	\mathbf v_{\rm D},
	\label{eq:dechirp_io_matrix}
\end{equation}
where the DAFT-domain equivalent channel matrix is
\begin{equation}
	{
		\mathbf H_{\rm D}
		=
		\mathbf{\Lambda}_{c_2}
		\mathbf F
		\left[
		\sum_{\ell=0}^{L-1}
		\breve h_\ell
		\mathbf D(\kappa_\ell)
		\mathbf{\Phi}_{{\rm D},\ell}
		\right]
		\mathbf F^{H}
		\mathbf{\Lambda}_{c_2}^{H}.
	}
	\label{eq:dechirp_channel_matrix}
\end{equation}

For completeness, define
\begin{equation}
	\mathbf K_{{\rm D},\ell}
	\triangleq
	\mathbf F
	\mathbf D(\kappa_\ell)
	\mathbf{\Phi}_{{\rm D},\ell}
	\mathbf F^{H}.
	\label{eq:dechirp_kernel_matrix}
\end{equation}
The $(m,q)$-th entry of the equivalent channel can then be expressed as
\begin{equation}
	H_{\rm D}[m,q]
	=
	e^{j2\pi c_2(q^2-m^2)}
	\sum_{\ell=0}^{L-1}
	\breve h_\ell
	K_{{\rm D},\ell}[m,q].
	\label{eq:dechirp_channel_element}
\end{equation}
The kernel
$\mathbf K_{{\rm D},\ell}$ in
\eqref{eq:dechirp_kernel_matrix} explicitly retains the effects of the
general transmit pulse and receive filter.

For comparison, consider the conventional chirp-dependent receive filter
in \cite{mirabella2026continuous}. At the sampling instant
$t_n=t_0+nT_s$, define
\begin{align}
	\left[
	\mathbf{\Phi}_{{\rm A},\ell}
	\right]_{n,k}
	&\triangleq
	\int_{-\infty}^{\infty}
	u_{\rm A}^{*}(-\lambda)
	p(t_n-\tau_\ell-kT_s-\lambda)
	\nonumber\\
	&\quad\times
	e^{-j2\pi[
		\nu_\ell+2\bar c_1(t_n-\tau_\ell)]
		\lambda}
	{\rm d}\lambda .
	\label{eq:conv_pulse_matrix}
\end{align}
After the $c_1$-dependent phase compensation in the conventional DAFT
demodulator, the corresponding DAFT-domain equivalent channel can be
written as
\begin{equation}
	\mathbf H_{\rm A}
	=
	\mathbf{\Lambda}_{c_2}
	\mathbf F
	\left[
	\sum_{\ell=0}^{L-1}
	\breve h_\ell
	\mathbf D(\kappa_\ell)
	\mathbf{\Phi}_{{\rm A},\ell}
	\right]
	\mathbf F^{H}
	\mathbf{\Lambda}_{c_2}^{H}.
	\label{eq:conventional_channel_matrix}
\end{equation}

The key difference between
\eqref{eq:dechirp_channel_matrix} and
\eqref{eq:conventional_channel_matrix} lies in their pulse-filtering
matrices. For the proposed scheme,
$\mathbf{\Phi}_{{\rm D},\ell}$ is generated by a fixed effective pulse
$\psi_\ell(t)$ and depends on the sample indices only through
$n-k$. In contrast, $\mathbf{\Phi}_{{\rm A},\ell}$ contains the explicit
sampling-index-dependent frequency term
\begin{equation}
	\nu_\ell
	+
	2\bar c_1(t_0+nT_s-\tau_\ell),
	\label{eq:conv_time_varying_frequency}
\end{equation}
which varies over an AFDM block. Therefore, a nonuniform response of the
conventional receive filter over the chirp-induced frequency sweep results
in additional time-varying weighting and may cause DAFT-domain channel
spreading.

\begin{figure}
	\centering
	\includegraphics[width=0.48\textwidth]{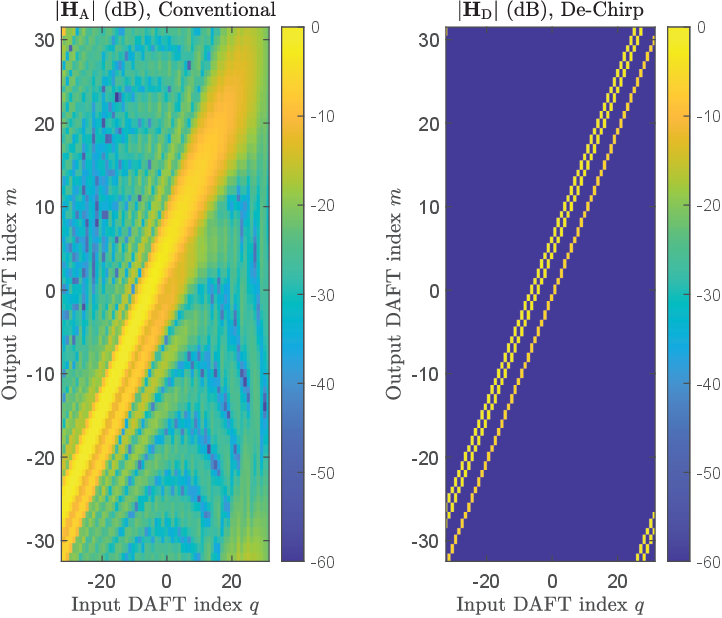}
	\caption{DAFT-domain equivalent channel matrices of the conventional and proposed de-chirp schemes with truncated RRC pulse shaping and filtering.}
	\label{comeq}
\end{figure}

To illustrate this difference under practical filtering, given $N=64$
and $L=3$, Fig.~\ref{comeq} compares the DAFT-domain equivalent channels
of the conventional and proposed schemes when a truncated RRC pulse is
employed at the transmitter and an RRC filter is used at the receiver.
The conventional scheme exhibits pronounced off-diagonal spreading
because the practical RRC filter cannot maintain a constant response over
the entire chirp-induced frequency sweep. By contrast, the proposed
de-chirp scheme retains a more concentrated equivalent-channel structure
because its receive filtering is time invariant after de-chirp.

To suppress such distortion, the conventional scheme \cite{mirabella2026continuous} requires a
sufficiently wide and approximately flat receive-filter response over the
chirp-induced frequency sweep. Such an ideal wide-and-flat response is
difficult to realize using a practical finite-duration filter. In
contrast, the proposed pre-filter de-chirp operation converts the
chirp-induced time variation into a fixed effective Doppler shift before
receive filtering, so that the subsequent LTI filtering does not introduce
the additional sampling-index-dependent weighting.

\section{Noise Power Analysis}
\label{subsec:noise_analysis}

We analyze the filtered noise and its equivalent power when noise
correlation is neglected in detection. Let $w(t)$ denote proper complex
AWGN with two-sided power spectral density (PSD) $N_0$. For the
conventional filter in \eqref{eq:conv_rx_filter}, define
$G_{\rm A}(f)=\mathcal F\{g_{\rm A}(t)\}$. In the proposed scheme,
de-chirp preserves the white Gaussian statistics before filtering.
The filtered-noise PSDs are therefore
\begin{equation}
	S_{\rm A}(f)=N_0|G_{\rm A}(f)|^2,\qquad
	S_{\rm D}(f)=N_0|G_{\rm D}(f)|^2.
	\label{eq:noise_output_psds}
\end{equation}
The filtered noise is generally colored. Orthogonal receive-filter
shifts spaced by $T_s$ yield uncorrelated samples, as for an ideal
rectangular actual frequency response of width $1/T_s$.
For general filters, we neglect noise correlation to simplify detection
and use an equivalent white-noise model with the same average power.
Filtering generally colors the continuous-time noise.
For $X\in\{\mathrm A,\mathrm D\}$, the autocorrelation of the noise
sampled at intervals of $T_s$ is
\begin{equation}
	R_X[k]=\int_{-\infty}^{\infty}
	S_X(f)e^{j2\pi f kT_s}\,{\rm d}f.
	\label{eq:noise_sample_corr}
\end{equation}
The samples are uncorrelated if $R_X[k]=0$ for $k\ne0$, as with
receive-filter shifts orthogonal at spacing $T_s$. In general this
condition need not hold. To avoid processing the correlation, we model
the noise as white with the same average power, $R_X[0]$.
Integrating the PSDs accounts for all noise passed by each filter and
gives the equivalent variances
\begin{align}
	\sigma_{\rm A}^2
	&=\int_{-\infty}^{\infty}S_{\rm A}(f)\,{\rm d}f
	=N_0\int_{-\infty}^{\infty}|U_{\rm A}(f)|^2\,{\rm d}f,
	\label{eq:noise_A}\\
	\sigma_{\rm D}^2
	&=\int_{-\infty}^{\infty}S_{\rm D}(f)\,{\rm d}f
	=N_0\int_{-\infty}^{\infty}|G_{\rm D}(f)|^2\,{\rm d}f.
	\label{eq:noise_D}
\end{align}
Equation~\eqref{eq:noise_A} follows from Parseval's identity: the chirp
factor preserves filter energy, although generally
$|G_{\rm A}(f)|\ne|U_{\rm A}(f)|$. The unitary DAFT preserves total
noise power, so these are also the average variances per DAFT component.
The white-noise model does not capture correlation effects in a
particular detection statistic.
Equal-energy filters give equal noise powers; an SNR comparison must
also account for their desired-signal gains.

For the bandwidth comparison, we use the ideal subcarrier model in
\eqref{eq:conv_filter_derivation} and rectangular responses
$U_{\rm A}^*(f)$ and $G_{\rm D}(f)$ with the same passband gain.
Recall that
\begin{equation}
	\kappa_\ell=\frac{\nu_\ell}{\Delta f}
	+2Nc_1\frac{t_0-\tau_\ell}{T_s},
	\label{eq:kappa_noise_analysis}
\end{equation}
and $\mu_\ell=\kappa_\ell\Delta f$. After de-chirp, the $q$-th
subcarrier on path $\ell$ has the fixed frequency
$
	f_{{\rm D},q,\ell}=(q+\kappa_\ell)\Delta f.
	\label{eq:dechirp_filter_frequency_noise}
$
The corresponding frequency argument of $U_{\rm A}^*(f)$ is
\begin{equation}
	(q+\kappa_\ell+2Nc_1n)\Delta f,\qquad n=0,\ldots,N-1.
	\label{eq:conv_filter_frequency_noise}
\end{equation}
Thus, the conventional response must also cover the chirp-induced
frequency sweep. For known path parameters, symmetric zero-centered
passbands with a half-subcarrier margin at each edge have the benchmark
widths
\begin{align}
	B_{\rm D}
	&=2\Delta f\left[
	\max_{\substack{q\in\mathcal Q\\\ell}}|q+\kappa_\ell|
	+\frac12\right],
	\label{eq:BD_noise}\\
	B_{\rm A}
	&=2\Delta f\left[
	\max_{\substack{q\in\mathcal Q,\,\ell\\0\le n\le N-1}}
	|q+\kappa_\ell+2Nc_1n|+\frac12\right].
	\label{eq:BA_noise}
\end{align}
Here, $B_{\rm A}$ is the width of $U_{\rm A}$, rather than the actual
response $G_{\rm A}$. With common passband gain $G_0$,
\eqref{eq:noise_A} and \eqref{eq:noise_D} become
\begin{equation}
	\sigma_{\rm A}^2=N_0|G_0|^2B_{\rm A},\qquad
	\sigma_{\rm D}^2=N_0|G_0|^2B_{\rm D}.
\end{equation}
For the same desired-signal power, the equivalent SNR gain under the
white-noise model is
\begin{equation}
	G_{\rm SNR}
	=10\log_{10}\frac{\sigma_{\rm A}^2}{\sigma_{\rm D}^2}
	=10\log_{10}\frac{B_{\rm A}}{B_{\rm D}}.
	\label{eq:SNR_gain}
\end{equation}

When only $\tau_{\max}$ and
$\nu_{\max}$ are known, let
$[\kappa_{\min},\kappa_{\max}]$ denote the resulting range in
\eqref{eq:kappa_noise_analysis}. Conservative bandwidths for this
benchmark are
\begin{align}
	B_{\rm D}^{\rm p}
	&=2\Delta f\left[
	\max_{\substack{q\in\mathcal Q\\
			\kappa\in[\kappa_{\min},\kappa_{\max}]}}
	|q+\kappa|+\frac12\right],\\
	B_{\rm A}^{\rm p}
	&=2\Delta f\left[
	\max_{\substack{q\in\mathcal Q,\,
			\kappa\in[\kappa_{\min},\kappa_{\max}]\\0\le n\le N-1}}
	|q+\kappa+2Nc_1n|+\frac12\right].
\end{align}

\section{Simulation Results}
In this section, we evaluate the proposed de-chirp AFDM scheme through numerical simulations. We adopt the RRC pulse-shaping scheme with wideband receive filtering in \cite{mirabella2026continuous} as the baseline. The chirp parameters $c_1$ and $c_2$ are selected
according to the design criteria in \cite{99}, where
$
c_1=\frac{2\nu_{\max}/\Delta f+2\xi_{\nu}+1}{2N}
$ with $\xi_{\nu}$ denoting the guard width,
and
$
c_2=\frac{1}{N^2\pi}.
$

\begin{figure*}[t]
	\centering
	\begin{minipage}[c]{0.32\linewidth} 
		\centering
		\includegraphics[width=1\textwidth]{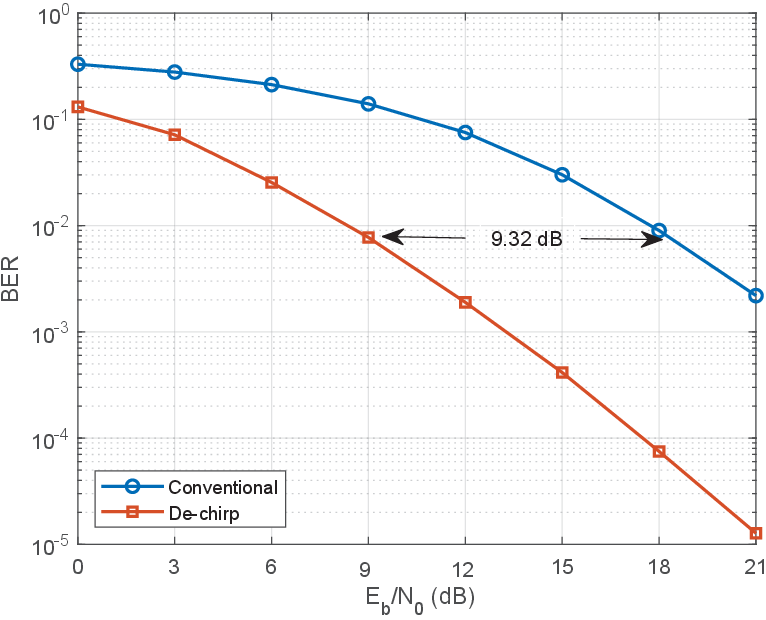}
		\caption{BER performance under OAMP detection and verification of the theoretical SNR gain}
		\label{ml}
	\end{minipage}
	\begin{minipage}[c]{0.32\linewidth}
		\centering
		\includegraphics[width=1\textwidth]{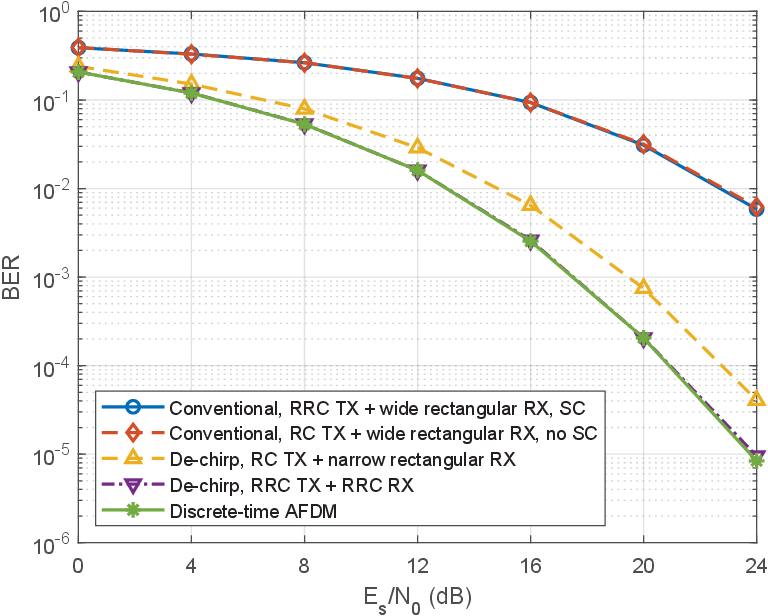}
		\caption{BER performance under MMSE detection with integer delays.}
		\label{mmseind}
	\end{minipage}
	\begin{minipage}[c]{0.32\linewidth} 
		\centering
		\includegraphics[width=1\textwidth]{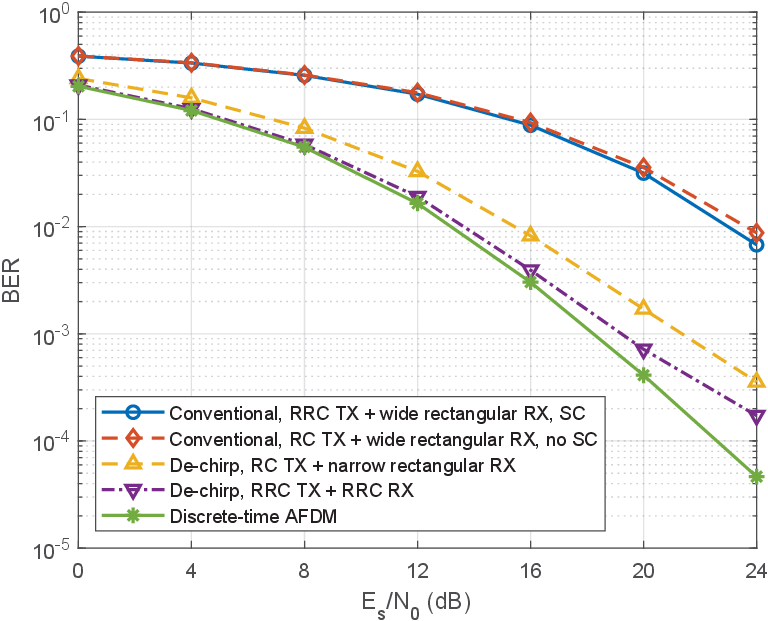}
		\caption{BER performance under MMSE detection with fractional delays.}
		\label{mmsefrd}
	\end{minipage}
\end{figure*}

We evaluate the BER performance of the proposed de-chirp AFDM scheme using orthogonal approximate message passing (OAMP) detection
with perfect CSI. To isolate the noise-bandwidth gain, we consider an
ideal path-aware benchmark with rectangular receive filters and the
half-bin guard used in the simulation. We set $N=64$, $t_0=0$ and 4-QAM symbols.
The three-path channel has delay indices
$\boldsymbol{\tau}/T_s=[0,1,2]$ and Doppler indices
$\boldsymbol{\nu}/\Delta f=[0,1,2]$. 
For these parameters, the receive-filter bandwidths are
$B_{\rm D}=81\Delta f$ and $B_{\rm A}=693\Delta f$. The predicted
horizontal BER separation is therefore
\begin{equation}
	G_{\rm SNR}
	=10\log_{10}\!\left(\frac{693}{81}\right)
	\approx 9.32~{\rm dB}.
\end{equation}
As shown in Fig.~\ref{ml}, the de-chirp AFDM scheme provides an
approximately $9.32$-dB SNR advantage at a given BER, consistent with
the predicted noise-bandwidth gain. The sampled codeword-difference
rank test yields rank three for the tested differences, consistent
with three-path diversity.

Next, we compare the BER performance of different schemes under minimum
mean-square error (MMSE)
detection. We set $N=256$, with the maximum delay and Doppler indices
both equal to $2$, and consider $L=4$. We first focus
on the case without fractional delays. Specifically, the delay index of
the first path is fixed at zero, while those of the remaining paths are
randomly selected as integers from $[0,2]$. The Doppler indices are
randomly generated within $[-2,2]$ and are allowed to take fractional
values. As shown in Fig.~\ref{mmseind}, the proposed de-chirp scheme maintains excellent BER
performance and significantly outperforms the baseline scheme, while
achieving performance close to that of ideal discrete-time AFDM. In
addition, when the de-chirp scheme employs the RRC pulse together with
the rectangular receive filter adopted in \cite{mirabella2026continuous}, a noticeable
performance loss is observed. This is because the rectangular receive
filter is not matched to the RRC transmit pulse, resulting in an SNR
penalty compared with RRC matched filtering.

Finally, we evaluate the BER performance of the considered schemes in the
presence of fractional delays. The simulation parameters are the same as
those used in Fig.~4, except that the delays of all paths other than the
first one are allowed to take fractional values within $[0,2]$.
As shown in Fig.~5, the proposed de-chirp scheme exhibits a certain
performance loss relative to ideal discrete-time AFDM when fractional
delays are present. This degradation is mainly caused by the residual
interference introduced by the misalignment between the fractionally
delayed pulses and the receive filter. Nevertheless, even under
fractional-delay channels, the proposed de-chirp scheme still
significantly outperforms the baseline scheme, demonstrating its effectiveness under fractional-delay channels.

\section{Conclusion}
In this paper, we proposed a CT AFDM reception scheme based on
pre-filter de-chirp for doubly selective channels. By characterizing the
coupling between the de-chirp operation and the channel, an equivalent signal representation with delay-coupled effective
Doppler shifts was established. Under general transmit pulse shaping and
LTI receive filtering, the resulting effective filtering kernel depends
on the sample indices only through their relative offset, thereby avoiding
the additional sampling-dependent weighting introduced by conventional
chirp-dependent filtering.
Furthermore, the receive-noise power was analyzed, and an explicit
expression for the SNR gap between the proposed and conventional schemes
was derived in terms of their required receive-filter bandwidths.
Simulation results verified the theoretical analysis and demonstrated that
the proposed scheme significantly outperforms the conventional
CT AFDM scheme while approaching the performance of ideal
discrete-time AFDM.

\bibliographystyle{IEEEtran}
\bibliography{AFDM}
\vfill
\end{document}